\documentclass[english,aip,pop,reprint]{revtex4-1}
\usepackage{ae,aecompl}

\usepackage[T1]{fontenc}
\usepackage[latin9]{inputenc}
\usepackage{amsmath}
\usepackage{amssymb}
\usepackage{graphicx}

\makeatletter

\providecommand{\tabularnewline}{\\}

\makeatother

\usepackage{babel}
\begin{document}
\title{Scalings Pertaining to Current Sheet Disruption Mediated by the Plasmoid
Instability}
\author{Yi-Min Huang}
\email{yiminh@princeton.edu}

\affiliation{Department of Astrophysical Sciences, Princeton Plasma Physics Laboratory,
Princeton Center for Heliophysics, and Max Planck/Princeton Center
for Plasma Physics, Princeton University, Princeton, New Jersey 08543,
USA}
\author{Luca Comisso}
\affiliation{Department of Astronomy and Columbia Astrophysics Laboratory, Columbia
University, New York, New York 10027, USA }
\author{Amitava Bhattacharjee}
\affiliation{Department of Astrophysical Sciences, Princeton Plasma Physics Laboratory,
Princeton Center for Heliophysics, and Max Planck/Princeton Center
for Plasma Physics, Princeton University, Princeton, New Jersey 08543,
USA}
\affiliation{Center for Computational Astrophysics, Flatiron Institute, New York,
NY 10010}
\begin{abstract}
Analytic scaling relations are derived for a phenomenological model
of the plasmoid instability in an evolving current sheet, including
the effects of reconnection outflow. Two scenarios are considered,
where the plasmoid instability can be triggered either by an injected
initial perturbation or by the natural noise of the system (here referred
to as the system noise). The two scenarios lead to different scaling
relations because the initial noise decays when the linear growth
of the plasmoid instability is not sufficiently fast to overcome the
advection loss caused by the reconnection outflow, whereas the system
noise represents the lowest level of fluctuations in the system. The
leading order approximation for the current sheet width at disruption
takes the form of a power law multiplied by a logarithmic factor,
and from that, the scaling relations for the wavenumber and the linear
growth rate of the dominant mode are obtained. When the effects of
the outflow are neglected, the scaling relations agree, up to the
leading order approximation, with previously derived scaling relations
based on a principle of least time. The analytic scaling relations
are validated with numerical solutions of the model.
\end{abstract}
\maketitle

\section{Introduction}

Thin current sheets, which appear to be ubiquitous in laboratory,
space, and astrophysical plasmas, are known to be unstable to the
plasmoid instability,\citep{Biskamp1986,LoureiroSC2007,BhattacharjeeHYR2009}
which disrupts current sheets to form smaller structures such as plasmoids
(or flux ropes) and secondary current sheets. Plasmoid-instability-mediated
disruption of reconnecting current sheets plays a crucial role in
triggering the transition from slow to fast magnetic reconnection,\citep{BhattacharjeeHYR2009,DaughtonRAKYB2009,CassakSD2009,HuangB2010,ShepherdC2010,HuangBS2011,HuangCB2017}
and also modifies the energy cascade in magnetohydrodynamic (MHD)
turbulence, making the energy power spectrum steeper than predicted
by traditional MHD turbulence theories.\citep{CarboneVM1990,HuangB2016,MalletSC2017,LoureiroB2017,BoldyrevL2017,ComissoHLHB2018,DongWHCB2018,WalkerBL2018}

In early theoretical studies of the plasmoid instability, it was commonly
assumed that the aspect ratio of the current sheet follows the Sweet-Parker
scaling $L/a\simeq\sqrt{S}$,\citep{Parker1957,Sweet1958} where $S\equiv LV_{A}/\eta$
is the Lundquist number. Here, $L$ and $a$ denote the half-length
and the half-width of the current sheet, respectively, $V_{A}$ is
the Alfv\'en speed, and $\eta$ is the magnetic diffusivity. For
resistive magnetohydrodynamics (MHD) model, this assumption yields
the scaling relations $\gamma\sim S^{1/4}$ for the linear growth
rate and $k\sim S^{3/8}$ for the wavenumber of the fastest growing
mode.\citep{TajimaS1997,LoureiroSC2007,BhattacharjeeHYR2009} Following
the same assumption, scaling relations have also been derived for
other models, including Hall MHD\citep{BaalrudBHG2011} and visco-resistive
MHD.\citep{ComissoG2016}

However, a current sheet in nature typically forms dynamically, starting
from a broader sheet that gradually thins down. The fact that the
Sweet-Parker-based scaling $\gamma\sim S^{1/4}$ diverges in the asymptotic
limit of $S\to\infty$ indicates that for a high-$S$ current sheet,
the disruption may occur before the current sheet realizes the Sweet-Parker
aspect ratio.\citep{PucciV2014} The precise conditions for current
sheet disruption have been a topic of active research in recent literature.\citep{PucciV2014,TeneraniVRP2015,TeneraniVPLR2016,UzdenskyL2016,ComissoLHB2016,ComissoLHB2017,HuangCB2017,TolmanLU2018}
By using a principle of least time, Comisso \emph{et al.} showed that
the current sheet width, the linear growth rate, and the dominant
mode wavenumber at disruption do not follow power-law scaling relations.\citep{ComissoLHB2016,ComissoLHB2017}
Huang \emph{et al.} proposed a phenomenological model incorporating
the effects of reconnection outflow.\citep{HuangCB2017} Numerical
solutions of this model have been extensively tested with results
from direct numerical simulations.

The objective of this work is to develop a methodology to obtain analytic
scaling relations for the phenomenological model of Huang \emph{et
al.} for a special class of evolving current sheets, which is modeled
by a Harris sheet thinning down exponentially in time, where the upstream
magnetic field remains constant. This paper is organized as follows.
Section \ref{sec:Phenomenological-Model} gives an overview of the
model. Section \ref{sec:Analytic-Solution-of} gives a formal solution
of the model equation based on the method of characteristics. The
solution is then applied in Section \ref{sec:Scaling-Relations} to
obtain scaling relations for key features at the disruption time,
such as the current sheet width, the linear growth rate, and the dominant
mode wavenumber. Here, we consider two possible scenarios. In the
first scenario, an initial noise is injected into the system to trigger
the plasmoid instability. In the second scenario, the plasmoid instability
evolves from the natural noise of the system, referred to as the ``system
noise'' in this paper.\footnote{It always requires some \textquotedblleft seed\textquotedblright{}
noise to trigger the plasmoid instability. In MHD simulations that
tend to be clean (i.e., with very low system noise due to discretization
and round-off errors), a seed noise is usually provided by adding
a random noise in the initial condition; this is an example of initial
noise. In contrast, particle-in-cell simulations typically do not
require an initial noise to trigger the plasmoid instability because
they are inherently noisy, but an initial noise can be seeded as well.
The random noise in particle-in-cell simulations is an example of
system noise. The same applies to laboratory and natural systems.
The plasmoid instability in such systems is usually triggered by the
system noise, but an initial noise may also be introduced in laboratory
experiments.} The two scenarios lead to different scaling relations because the
initial noise tends to decay during an early time when the linear
growth of the plasmoid instability is not sufficiently fast to overcome
the advection loss caused by the reconnection outflow; on the other
hand, the system noise represents the lowest level of fluctuations
pertaining to the system even when no external noise is injected.
In Section \ref{sec:Discussions}, we consider the case where the
effects of the outflow are neglected. In this case, we show that the
scaling relations agree with that of Comisso \emph{et al. }up to the
leading order approximation. We also validate the analytic scaling
relations by comparing them with numerical solutions. We conclude
in Section \ref{sec:Conclusion}.

\section{Phenomenological Model\label{sec:Phenomenological-Model}}

\begin{figure}

\includegraphics[scale=0.43]{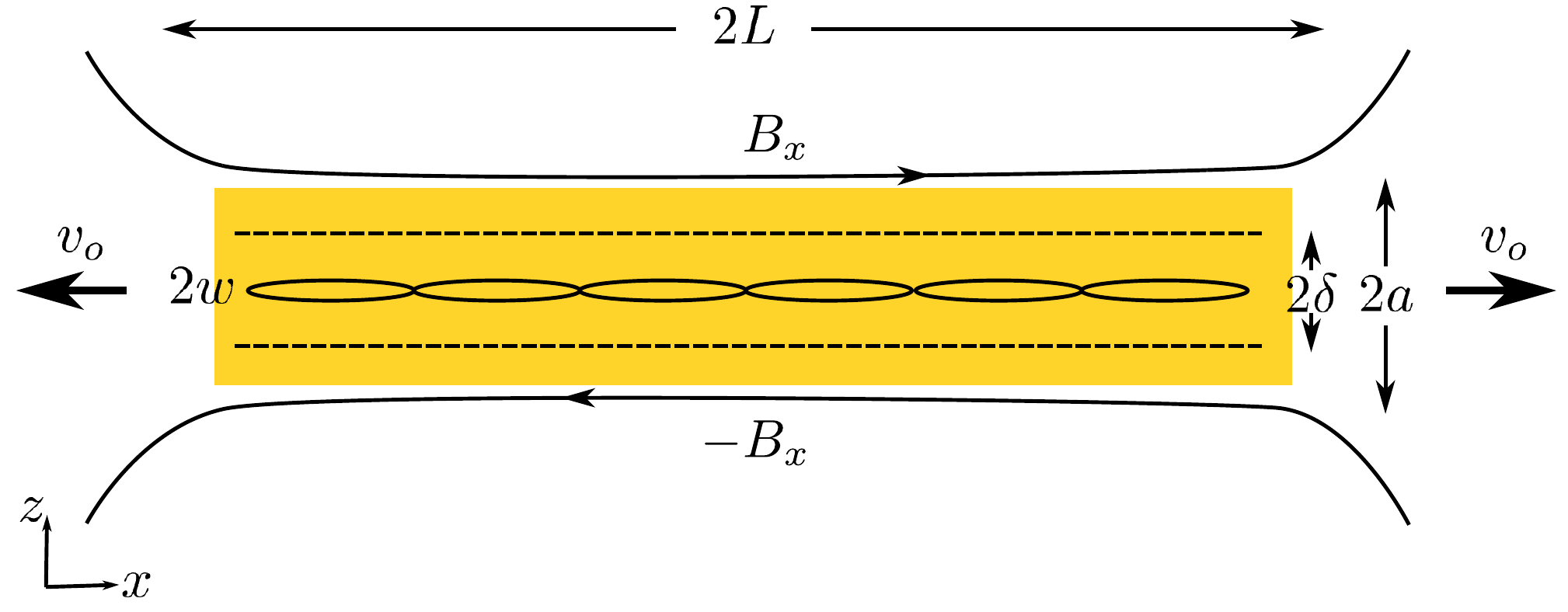}

\caption{Schematic diagram of the plasmoid instability in a reconnecting current
sheet. Here, the shaded area is the current sheet of a length $2L$
and a width $2a$. The length and width can both be functions of time.
The magnetic fields immediately outside the current sheet are $\pm B_{x}$,
and the reconnection outflow speed is $v_{o}$. Within the current
sheet are two additional length scales: the inner layer width $2\delta$,
and the magnetic island width $2w$. Current sheet disruption occurs
when the magnetic island width exceeds the inner layer width. \label{fig:Schematic-of-plasmoid}}

\end{figure}

In this Section, we give a brief outline of the phenomenological model.
The reader is referred to Ref.~{[}\onlinecite{HuangCB2017}{]} for
the details of the derivation. Figure \ref{fig:Schematic-of-plasmoid}
shows a schematic diagram for the system we consider here. The following
equation governs the linear phase of the plasmoid instability in an
evolving current sheet:

\begin{equation}
\partial_{t}f-k\frac{v_{o}}{L}\partial_{k}f=\left(\gamma-\frac{v_{o}}{2L}+\frac{1}{2L}\frac{dL}{dt}\right)f.\label{eq:model}
\end{equation}
Here, $f(k,t)\equiv|\hat{B}_{z}(k,t)|/B_{0}L_{0}$ is the Fourier
amplitude of the magnetic field fluctuation normal to the current
sheet as a function of the wavenumber $k$ and the time $t$, normalized
to the characteristic magnetic field $B_{0}$ and length $L_{0}$
of the system, $v_{o}$ is the outflow speed, $\gamma(k,t)$ is the
linear growth rate of the tearing instability, and $L$ is the half-length
of the current sheet. Here, the linear growth rate $\gamma$ explicitly
depends on time because the current sheet can evolve in time. The
differential operator in the left-hand-side incorporates the stretching
effect on wavelength due to the reconnection outflow; the right-hand-side
represents the effects of linear growth, advection loss due to the
outflow, and evolution of the current sheet length. The domain of
wavenumber $k$ is limited to $k\ge\pi/L$ because the wavelength
must be smaller than the current sheet length $2L$.

The linear growth rate $\gamma(k,t)$ depends on the profile of magnetic
field $B_{x}(z,t)$, which can be a function of time depending on
the system under consideration. The model also allows the current
sheet half-length $L$ to change in time, as represented by the $(1/2L)dL/dt$
term. The outflow speed $v_{0}$ can also be time-dependent. To integrate
the model equation, we need $\gamma(k,t)$, $v_{o}(t)$, $L(t)$,
and an initial condition $f=f_{0}(k)$ as inputs. With these conditions
provided, we can integrate equation (\ref{eq:model}) in time until
the plasmoid instability enters the nonlinear regime, which occurs
when the typical half-width $w$ of plasmoids exceeds the inner layer
half-width $\delta$. At that moment, the fluctuating part of the
current density $\tilde{J}$ is of the same order of the background
current density $J$, implying that the current sheet has lost its
integrity. Therefore, we take $w=\delta$ as the condition for current
sheet disruption.

This condition for current sheet disruption depends critically on
the half-width $w$ of plasmoids. However, the precise definition
of $w$ is nuanced. The nuance comes from the fact that while $w$
is well-defined when the magnetic fluctuation $\tilde{B}$ is a single-mode
Fourier harmonic (as is often assumed in tearing mode analyses), it
is not so when the fluctuation is a superposition of a continuum of
Fourier modes. In this model problem, fortunately, a meaningful half-width
$w$ of plasmoids can be defined because, as we will see, the solution
$f$ typically becomes localized around a dominant wavenumber $k_{d}$
as time progresses. For a spectrum $f(k)$ localized around $k=k_{d}$,
we define the plasmoid half-width $w$ as

\begin{equation}
w=2\sqrt{\frac{\tilde{B}}{k_{d}B_{x}'}},\label{eq:island_width}
\end{equation}
where the prime denotes $d/dz$ evaluated at $z=0$. In Eq.~(\ref{eq:island_width}),
the fluctuation amplitude $\tilde{B}$ must take into account the
contribution from all the neighboring modes of the dominant mode.
Precisely, a superposition of all the modes in the range $k\in[k_{d}/\xi,k_{d}\xi]$
gives a fluctuation amplitude 

\begin{equation}
\tilde{B}=\left(\frac{L_{0}^{2}}{\pi L}\int_{k_{d}/\xi}^{k_{d}\xi}f(k')^{2}dk'\right)^{1/2}B_{0}.\label{eq:amplitude}
\end{equation}
Here, the $O(1)$ parameter $\xi$ sets the range of superposition.
As the Fourier spectrum $f(k)$ is usually well-localized at the disruption
time, the amplitude $\tilde{B}$ is insensitive to the choice of $\xi$,
provided that $\xi$ is not too close to unity (e.g., $\xi=1.5$ was
used in Ref.~{[}\onlinecite{HuangCB2017}{]}). Note that for a single-mode
perturbation, the definition (\ref{eq:island_width}) reduces to the
usual expression for magnetic island half-width.\citep{Biskamp1993}

The plasmoid half-width $w$ must be compared with the inner layer
half-width $\delta$ to determine the condition for current sheet
disruption. As the inner layer half-width $\delta$ depends on the
wavenumber, we adopt the inner layer half-width at the dominant wavenumber
$k_{d}$ for this comparison. For resistive tearing modes we consider
in this study, the inner layer half-width for $k=k_{d}$ is given
by
\begin{equation}
\delta=\left(\frac{\eta\,\gamma(k_{d})}{\left(k_{d}V_{A}\right)'^{2}}\right)^{1/4},\label{eq:inner_general}
\end{equation}
where $\eta$ is the magnetic diffusivity. Here, the prime again denotes
$d/dz$ evaluated at $z=0$, and $V_{A}$ is the Alfv\'en speed of
the reconnecting component $B_{x}$. The criterion $w=\delta$, with
$w$ and $\delta$ defined by Eqs.~(\ref{eq:island_width}--\ref{eq:inner_general}),
then determines the condition for current sheet disruption. This criterion
has been extensively tested and validated with numerical simulations.\citep{HuangCB2017}

\section{Integral Solution of the Model Equation for a thinning Harris Sheet\label{sec:Analytic-Solution-of}}

Up to now, the model is general regarding the time evolution of both
the current sheet profile and the reconnection outflow. To fix ideas
and make the model analytically tractable, we assume that the current
sheet can be modeled by a Harris sheet and the time evolution of the
width $a$ obeys the form
\begin{equation}
a^{2}=a_{\infty}^{2}+(a_{0}^{2}-a_{\infty}^{2})\exp(-2t/\tau).\label{eq:a-solution}
\end{equation}
In this form, assuming $a_{\infty}\ll a_{0}$, the current sheet width
$a$ decreases exponentially at early times with $a\simeq a_{0}e^{-t/\tau}$,
and $a\to a_{\infty}$ as $t\to\infty$. We further assume that the
upstream magnetic field $B_{x}$ and the current sheet half-length
$L$ remain fixed in time and that the outflow speed $v_{o}$ is the
upstream Alfv\'en speed $V_{A}$. Under these conditions, if the
plasmoid instability does not disrupt the current sheet, eventually
the current sheet will approach the Sweet-Parker current sheet, which
is a steady-state solution of reconnection governed by resistive MHD.
Therefore, we set the asymptotic half-width $a_{\infty}$ to be the
Sweet-Parker width, i.e., $a_{\infty}=L/\sqrt{S}$, where $S\equiv LV_{A}/\eta$
is the Lundquist number. These assumptions are consistent with direct
numerical simulations reported in Ref.~{[}\onlinecite{HuangCB2017}{]}.
Heuristic derivations for Eq.~(\ref{eq:a-solution}) can be found
in Refs.~{[}\onlinecite{Kulsrud2005}{]} and {[}\onlinecite{HuangCB2017}{]}.

Under the assumption that the upstream $B_{x}$ and the current sheet
length $L$ are fixed in time, it is natural to choose $B_{0}=B_{x}$
and $L_{0}=L$ as the characteristic magnetic field and length scale.
Hereafter, we normalize lengths and times with respect to the current
sheet half-length $L$ and the Alfv\'en time $\tau_{A}\equiv L/V_{A}$,
as follows: $\hat{t}=t/\tau_{A}$, $\hat{\tau}=\tau/\tau_{A}$, $\hat{k}=kL$,
$\hat{a}=a/L$, and $\hat{\gamma}=\gamma\tau_{A}$. The model equation
in normalized variables is
\begin{equation}
\partial_{\hat{t}}f-\hat{k}\partial_{\hat{k}}f=\left(\hat{\gamma}-\frac{1}{2}\right)f,\label{eq:model-normalized}
\end{equation}
where $dL/dt=0$ is assumed.

We can solve Eq.~(\ref{eq:model-normalized}) by the method of characteristics.
The characteristic starting from $\hat{k}=\hat{k}_{0}$ at $\hat{t}=\hat{t}_{0}$
is given by 
\begin{equation}
\hat{k}\left(\hat{t};\hat{k}_{0},\hat{t}_{0}\right)=\hat{k}_{0}e^{-(\hat{t}-\hat{t}_{0})},\label{eq:characteristic}
\end{equation}
which represents the mode-stretching effect due to the reconnection
outflow. Now the differential operator on the left-hand-side of Eq.
(\ref{eq:model-normalized}) is simply the time derivative along characteristics
and hence the solution can be obtained by integrating along them,
yielding 
\begin{equation}
f(\hat{k},\hat{t})=f(\hat{k}_{0},\hat{t}_{0})\exp\left[\int_{\hat{t}_{0}}^{\hat{t}}\hat{\gamma}\left(\hat{k}(\hat{t}'),\hat{t}'\right)d\hat{t}'-\frac{\hat{t}-\hat{t}_{0}}{2}\right].\label{eq:solution-3}
\end{equation}

To perform the integration, we need the linear growth rate $\hat{\gamma}$
as a function of the wavenumber and time. The linear growth rate of
the tearing mode is governed by the tearing stability index $\Delta'$,\citep{FurthKR1963}
which depends entirely on the solution of linearized ideal MHD force-free
equation in the outer region away from the resonant surface (where
$\mathbf{k}\cdot\mathbf{B}=0$). The tearing instability requires
the condition $\Delta'>0$ being satisfied. While the complete dispersion
relation for resistive tearing modes can be expressed in terms of
$\Delta'$, the expression is involved and requires solving a transcendental
equation.\citep{CoppiGPRR1976} Analytically tractable approximations
can be obtained in two limits: the small-$\Delta'$ regime for short-wavelength
modes and the large-$\Delta'$ regime for long-wavelength modes. In
the small-$\Delta'$ regime (large $k$ and $\Delta'\delta\ll1$),
the linear growth rate is given by \citep{FurthKR1963}
\begin{equation}
\gamma_{s}=C_{\Gamma}\eta^{3/5}(\Delta'/2)^{4/5}\left(kV_{A}\right)'^{2/5},\label{eq:small-delta-prime}
\end{equation}
where $C_{\Gamma}=\left(\pi{}^{-1}\Gamma(1/4)/\Gamma(3/4)\right)^{4/5}\approx0.953$.
In the the large-$\Delta'$ regime (small $k$ and $\Delta'\delta\gg1$),
the linear growth rate is approximately given by \citep{CoppiGPRR1976}
\begin{equation}
\gamma_{l}=\eta^{1/3}\left(kV_{A}\right)'^{2/3}.\label{eq:large-delta-prime}
\end{equation}
Here, the derivative $\left(kV_{A}\right)'$ is evaluated at the resonant
surface. For a Harris sheet profile with $V_{A}(z)=V_{A}\tanh(z/a)$,
the tearing stability index $\Delta'$ is given by
\begin{equation}
\Delta'=\frac{2}{a}\left(\frac{1}{ka}-ka\right).\label{eq:delta_prime}
\end{equation}
From Eqs.~(\ref{eq:small-delta-prime}--\ref{eq:delta_prime}),
we obtain the normalized growth rates for a Harris sheet in the small-$\Delta'$
regime 
\begin{equation}
\hat{\gamma}_{s}=C_{\Gamma}\hat{a}^{-2}S^{-3/5}\hat{k}^{-2/5}(1-\hat{k}^{2}\hat{a}^{2})^{4/5}\label{eq:small-delta-prime-normalized}
\end{equation}
and in the large-$\Delta'$ regime
\begin{equation}
\hat{\gamma}_{l}=\hat{a}^{-2/3}S^{-1/3}\hat{k}^{2/3}.\label{eq:large-delta-prime-normalized}
\end{equation}
The two asymptotic limits (\ref{eq:small-delta-prime-normalized})
and (\ref{eq:large-delta-prime-normalized}) can be smoothly connected
by a function 
\begin{equation}
\hat{\gamma}=\frac{\hat{\gamma_{s}}\hat{\gamma_{l}}}{\left(\hat{\gamma_{s}}^{\zeta}+\hat{\gamma_{l}}^{\zeta}\right)^{1/\zeta}},\label{eq:uniform_approx}
\end{equation}
for an arbitrary $\zeta$. In this work, we take the value $\zeta=3/2$, which gives a nearly exact approximation of the true dispersion relation.
Furthermore, we will neglect the factor $(1-\hat{k}^{2}\hat{a}^{2})$
in Eq. (\ref{eq:small-delta-prime-normalized}) for our analytic derivations\emph{.
}This approximation can be justified\emph{ a posteriori }by noting
that the dominant mode wavenumber $\hat{k}_{d}$ follows the same
scaling as the fastest growing mode wavenumber $\hat{k}_{\text{max}}$
{[}see Eq. (\ref{eq:dominant_wavenumber}) and the discussion thereafter{]}
and that $\hat{k}_{\text{max}}\hat{a}\simeq(S\hat{a}){}^{-1/4}$ (see
below); therefore, in the high-$S$ regime, the condition $\hat{k}\hat{a}\ll1$
is satisfied in the neighborhood (in the $k$-space) of the dominant
mode, which is our primary interest. Note that, from the scaling of
the disruption current sheet width, Eq. (\ref{eq:ad}), the condition
$S\hat{a}\gg1$ is satisfied. Putting these approximations together
yields an approximate expression for the linear growth rate that is
valid for $\hat{k}\hat{a}\ll1$, i.e.,
\begin{equation}
\hat{\gamma}\simeq\frac{C_{\Gamma}S^{-3/5}\hat{k}^{-2/5}\hat{a}^{-2}}{\left(1+C_{\Gamma}^{3/2}S^{-2/5}\hat{k}^{-8/5}\hat{a}^{-2}\right)^{2/3}}.\label{eq:gamma}
\end{equation}
We will apply this expression to Eq.~(\ref{eq:solution-3}) to derive
scaling relations in Section \ref{sec:Scaling-Relations}.

The fastest growing mode occurs at the transition between the small-$\Delta'$
and the large-$\Delta'$ regimes, i.e., at $\hat{k}\hat{a}\simeq(S\hat{a}){}^{-1/4}$
where $\hat{\gamma}_{s}\simeq\hat{\gamma}_{l}$. More precisely, the
fastest growing wavenumber $\hat{k}_{\text{max}}$ is given by (see,
\emph{e.g.,} Ref.~{[}\onlinecite{Schindler2007}{]})

\begin{equation}
\hat{k}_{\text{max}}\approx1.358\,(S\hat{a}){}^{-1/4}\hat{a}^{-1}\label{eq:kmax}
\end{equation}
and the corresponding growth rate is
\begin{equation}
\hat{\gamma}_{\text{max}}\approx0.623\,S^{-1/2}\hat{a}^{-3/2}.\label{eq:gamma_max}
\end{equation}

\section{Scaling Relations for Current Sheet Disruption \label{sec:Scaling-Relations}}

In this Section we derive scaling relations for key features at current
sheet disruption in the high-$S$ regime, under the assumptions detailed
in Sec.~\ref{sec:Analytic-Solution-of}. Specifically, the disruption
time $\hat{t}_{d}$, the current sheet width $\hat{a}_{d}$, the dominant
mode wavenumber $\hat{k}_{d}$, and the linear growth rate $\hat{\gamma}_{d}$
of the dominant mode are derived analytically. Here we consider two
cases. In the first case, an initial noise is injected into the system.
The initial noise is assumed to be much larger than the system noise;
therefore the latter is ignored and set to zero in the analysis. In
the second case, there is no initial noise, and the system noise is
the seed from which the plasmoid instability develops.

\subsection{Case I --- Initial Perturbation\label{subsec:Case-I}}

Let the initial perturbation be $f_{0}(\hat{k}_{0})$ at $\hat{t}_{0}=0$.
To integrate the linear growth rate along a characteristic in Eq.~(\ref{eq:solution-3}),
it is convenient to define a new variable
\begin{equation}
\varphi\equiv C_{\Gamma}^{-3/4}S^{1/5}\hat{k}^{4/5}\hat{a}.\label{eq:integrate_var}
\end{equation}
In the high-$S$ regime we consider here, it can be shown \emph{a
posteriori} that current sheet disruption occurs when $\hat{a}_{d}\gg\hat{a}_{\infty}$,
because $\hat{a}_{d}$ follows a weaker scaling with respect to $S$
{[}see Eq.~(\ref{eq:ad}){]} than $\hat{a}_{\infty}$, which is assumed
to be the Sweet-Parker width $\hat{a}_{SP}=S^{-1/2}$. Hence, we will
ignore $a_{\infty}$ in Eq.~(\ref{eq:a-solution}) and assume $\hat{a}=\hat{a}_{0}e^{-\hat{t}/\hat{\tau}}$
in the following analysis. The combined effect of mode stretching
($\hat{k}=\hat{k}_{0}e^{-(\hat{t}-\hat{t}_{0})}$) and current sheet
thinning $(\hat{a}=\hat{a}_{0}e^{-\hat{t}/\hat{\tau}})$ implies that
the variable $\varphi$ decreases exponentially in time as 
\begin{equation}
\varphi=\varphi_{0}e^{-\hat{t}/\hat{\tau}_{*}},\label{eq:newvar}
\end{equation}
where the time scale $\hat{\tau}_{*}$ is defined as 
\begin{equation}
\hat{\tau}_{*}=\frac{5\hat{\tau}}{4\hat{\tau}+5}.\label{eq:hybrid_time}
\end{equation}
The linear growth rate, Eq.~(\ref{eq:gamma}), can be expressed in
terms of $\varphi$ instead of $\hat{k}$ as 
\begin{equation}
\hat{\gamma}\simeq C_{\Gamma}^{5/8}S^{-1/2}\hat{a}^{-3/2}\frac{\varphi{}^{-1/2}}{\left(1+\varphi^{-2}\right)^{2/3}}.\label{eq:growthrate_newvar}
\end{equation}

We can now integrate the growth rate along the characteristic with
a change of variable from $\hat{t}$ to $\varphi$, yielding
\begin{align}
 & \int_{0}^{\hat{t}}\hat{\gamma}(\hat{k}_{0}e^{-\hat{t}'},\hat{t}')d\hat{t}'\nonumber \\
\simeq & -C_{\Gamma}^{5/8}S^{-1/2}\hat{a}^{-3/2}\varphi^{3\hat{\tau}_{*}/2\hat{\tau}}\hat{\tau}_{*}\int_{\varphi_{0}}^{\varphi}\frac{\varphi'^{-3/2-3\hat{\tau}_{*}/2\hat{\tau}}}{\left(1+\varphi'^{-2}\right)^{2/3}}d\varphi'\nonumber \\
= & \,\frac{\hat{\tau}_{\sharp}}{2}C_{\Gamma}^{5/8}S^{-1/2}\hat{a}^{-3/2}\varphi^{3\hat{\tau}_{*}/2\hat{\tau}}\left.G(\varphi';\hat{\tau})\right|_{\varphi_{0}}^{\varphi},\label{eq:gamma_int_0}
\end{align}
where
\begin{equation}
G(\varphi';\hat{\tau})=\varphi'^{-1/2-3\hat{\tau}_{*}/2\hat{\tau}}\,_{2}F_{1}\left(\frac{\hat{\tau}_{*}}{\hat{\tau}_{\sharp}},\frac{2}{3};\frac{\hat{\tau}_{*}}{\hat{\tau}_{\sharp}}+1;-\varphi'^{-2}\right).\label{eq:G_function}
\end{equation}
Here, $_{2}F_{1}$ is the hypergeometric function\citep{AbramowitzS1972}
and 
\begin{equation}
\hat{\tau}_{\sharp}\equiv\frac{4\hat{\tau}\hat{\tau}_{*}}{\hat{\tau}+3\hat{\tau}_{*}}=\dfrac{5\hat{\tau}}{\hat{\tau}+5}.\label{eq:tau_palm}
\end{equation}

To proceed from Eq.~(\ref{eq:gamma_int_0}), we further make the
approximation that in the high-$S$ regime, the contribution from
the lower boundary, $G(\varphi_{0};\hat{\tau})$, is negligible compared
to the contribution from the upper boundary, $G(\varphi;\hat{\tau})$,
at the disruption time. This approximation can be justified \emph{a
posteriori} using the scaling relation of the disruption current sheet
width $\hat{a}_{d}$ in Eq.~(\ref{eq:ad}). Because $\hat{a}_{d}$
decreases as $S$ increases, it follows from the relation $\varphi/\varphi_{0}=(\hat{a}/\hat{a}_{0})^{\hat{\tau}/\hat{\tau}_{*}}$
that in the high-$S$ regime, the condition $\varphi\ll\varphi_{0}$
is satisfied at the disruption time, where the variable $\varphi$
is typically an $O(1)$ quantity in the neighborhood of the dominant
mode {[}see the discussion in the paragraph below Eq.~(\ref{eq:solution}){]}.
The function $G(\varphi';\hat{\tau})$ is monotonically decreasing
with respect to $\varphi'$. Its overall behavior is well captured
by two asymptotic expressions for small and large $\varphi'$, which
can be obtained using the asymptotic expansion of the hypergeometric
function near $x\to\infty$:
\begin{align}
_{2}F_{1}\left(c,b;c+1;-x\right)= & \frac{\Gamma(c+1)\Gamma(b-c)}{\Gamma(b)}x^{-c}\nonumber \\
 & +\frac{c}{c-b}x^{-b}+O\left(x^{-b-1}\right)\label{eq:asymptotic}
\end{align}
and the Taylor expansion near $x=0$:
\begin{equation}
_{2}F_{1}\left(c,b;c+1;-x\right)=1-\frac{bcx}{c+1}+O(x^{2});\label{eq:taylor}
\end{equation}
for the special case with $b=c$, Eq.~(\ref{eq:asymptotic}) is not
valid and must be replaced by
\begin{equation}
_{2}F_{1}\left(b,b;b+1;-x\right)=bx^{-b}\log(x)+O(x^{-b}).\label{eq:asymptotic_special}
\end{equation}
It follows that the function $G(\varphi';\hat{\tau})$ is approximately
\begin{align}
G(\varphi';\hat{\tau})\simeq & \frac{\Gamma\left(\frac{5\hat{\tau}+10}{4\hat{\tau}+5}\right)\Gamma\left(\frac{5\hat{\tau}-5}{12\hat{\tau}+15}\right)}{\Gamma(2/3)}\nonumber \\
 & +\frac{3\hat{\tau}+15}{5-5\hat{\tau}}\varphi'^{10(\hat{\tau}-1)/(12\hat{\tau}+15)}\label{eq:small}
\end{align}
when $\varphi'\lesssim O(1)$, and 
\begin{equation}
G(\varphi';\hat{\tau})\simeq\varphi'^{-(2\hat{\tau}+10)/(4\hat{\tau}+5)}\label{eq:large}
\end{equation}
when $\varphi'\gtrsim O(1)$. For the special case $\hat{\tau}=1$
that yields $\hat{\tau}_{*}/\hat{\tau}_{\sharp}=2/3$, Eq.~(\ref{eq:small})
is replaced by
\begin{equation}
G(\varphi';\hat{\tau})\simeq-\frac{4}{3}\log\varphi'.\label{eq:small_special}
\end{equation}
Because $\varphi_{0}\gg\varphi\sim O(1)$, it follows from Eqs.~(\ref{eq:small}
-- \ref{eq:small_special}) that $G(\varphi_{0};\hat{\tau})\ll G(\varphi;\hat{\tau}).$

Ignoring $G(\varphi_{0};\hat{\tau})$, we can simplify Eq.~(\ref{eq:gamma_int_0})
as
\begin{equation}
\int_{0}^{\hat{t}}\hat{\gamma}(\hat{k}_{0}e^{-\hat{t}'},\hat{t}')d\hat{t}'\simeq h\,g(\varphi;\hat{\tau}),\label{eq:gamma_int}
\end{equation}
where
\begin{equation}
h\equiv\frac{\hat{\tau}_{\sharp}}{2}C_{\Gamma}^{5/8}S^{-1/2}\hat{a}^{-3/2}\label{eq:h}
\end{equation}
and 
\begin{equation}
g(\varphi;\hat{\tau})\equiv\varphi^{-1/2}\,_{2}F_{1}\left(\frac{\hat{\tau}_{*}}{\hat{\tau}_{\sharp}},\frac{2}{3};\frac{\hat{\tau}_{*}}{\hat{\tau}_{\sharp}}+1;-\varphi{}^{-2}\right).\label{eq:g_phi}
\end{equation}
The solution of the model equation can now be written as 
\begin{equation}
f(\hat{k},\hat{t})\simeq f_{0}(\hat{k}_{0})\exp\left[h\,g(\varphi;\hat{\tau})-\frac{\hat{t}}{2}\right],\label{eq:solution}
\end{equation}
with $\hat{k}_{0}=\hat{k}e^{\hat{t}}$, $h=\hat{\tau}_{\sharp}C_{\Gamma}^{5/8}S^{-1/2}\hat{a}^{-3/2}/2$,
and $\varphi=C_{\Gamma}^{-3/4}S^{1/5}\hat{k}^{4/5}\hat{a}$ being
substituted into the right-hand-side.

The dominant wavenumber $\hat{k}_{d}$ at the disruption time can
be obtained by locating the maximum of $f$. Because the exponential
factor in Eq.~(\ref{eq:solution}) is highly localized, whereas most
random fluctuations have a broadband spectrum, we further assume that
$f_{0}(\hat{k}_{0})$ is slowly varying compared to the exponential
factor. Under this assumption, the dominant wavenumber is approximately
the one that maximizes the exponent in Eq.~(\ref{eq:solution}).\footnote{This assumption is not valid if $f_{0}(\hat{k}_{0})$ is narrowband.
In that case, we have to locate the maximum of the full $f(\hat{k},\hat{t})$.} Let $\varphi=\varphi_{d}$ correspond to the maximum of $g(\varphi;\hat{\tau})$.
The value of $\varphi_{d}$ depends on $\hat{\tau}$ and, in general,
has to be obtained numerically; typically, $\varphi_{d}$ is an $O(1)$
quantity. With the value of $\varphi_{d}$ obtained, the dominant
wavenumber $\hat{k}_{d}$ and linear growth rate $\hat{\gamma}_{d}$
are given by 
\begin{equation}
\hat{k}_{d}\simeq c_{k}\left(S\hat{a}_{d}\right)^{-1/4}\hat{a}_{d}^{-1}\label{eq:dominant_wavenumber}
\end{equation}
and 
\begin{equation}
\hat{\gamma}_{d}\simeq c_{\gamma}S^{-1/2}\hat{a}_{d}^{-3/2},\label{eq:dominant_gamma}
\end{equation}
where the constants $c_{k}$ and $c_{\gamma}$ are defined as 
\begin{equation}
c_{k}\equiv\left(\varphi_{d}C_{\Gamma}^{3/4}\right)^{5/4}\label{eq:ck}
\end{equation}
and 
\begin{equation}
c_{\gamma}\equiv C_{\Gamma}^{5/8}\frac{\varphi{}_{d}^{-1/2}}{\left(1+\varphi_{d}^{-2}\right)^{2/3}}.\label{eq:cgamma}
\end{equation}
Equation (\ref{eq:dominant_wavenumber}) shows that the dominant wavenumber
and the fastest growing wavenumber follow the same scaling {[}cf.~Eq.~(\ref{eq:kmax}){]}.
However, as we will see in Table \ref{tab:constants}, typically the
constant $c_{k}$ is smaller than the coefficient in Eq.~(\ref{eq:kmax}).
Therefore, the dominant wavenumber is smaller than the fastest-growing
wavenumber. This discrepancy between the dominant wavenumber and the
fastest-growing wavenumber is due to the fact that the mode amplitude
at a given wavenumber $\hat{k}$ is determined by the entire history
of the linear growth rate following the characteristic. Even though
the dominant mode is not the fastest growing mode at the moment, it
has higher linear growth rates at earlier times. \citep{HuangCB2017}

\begin{figure}
\begin{centering}
\includegraphics[width=3.5in]{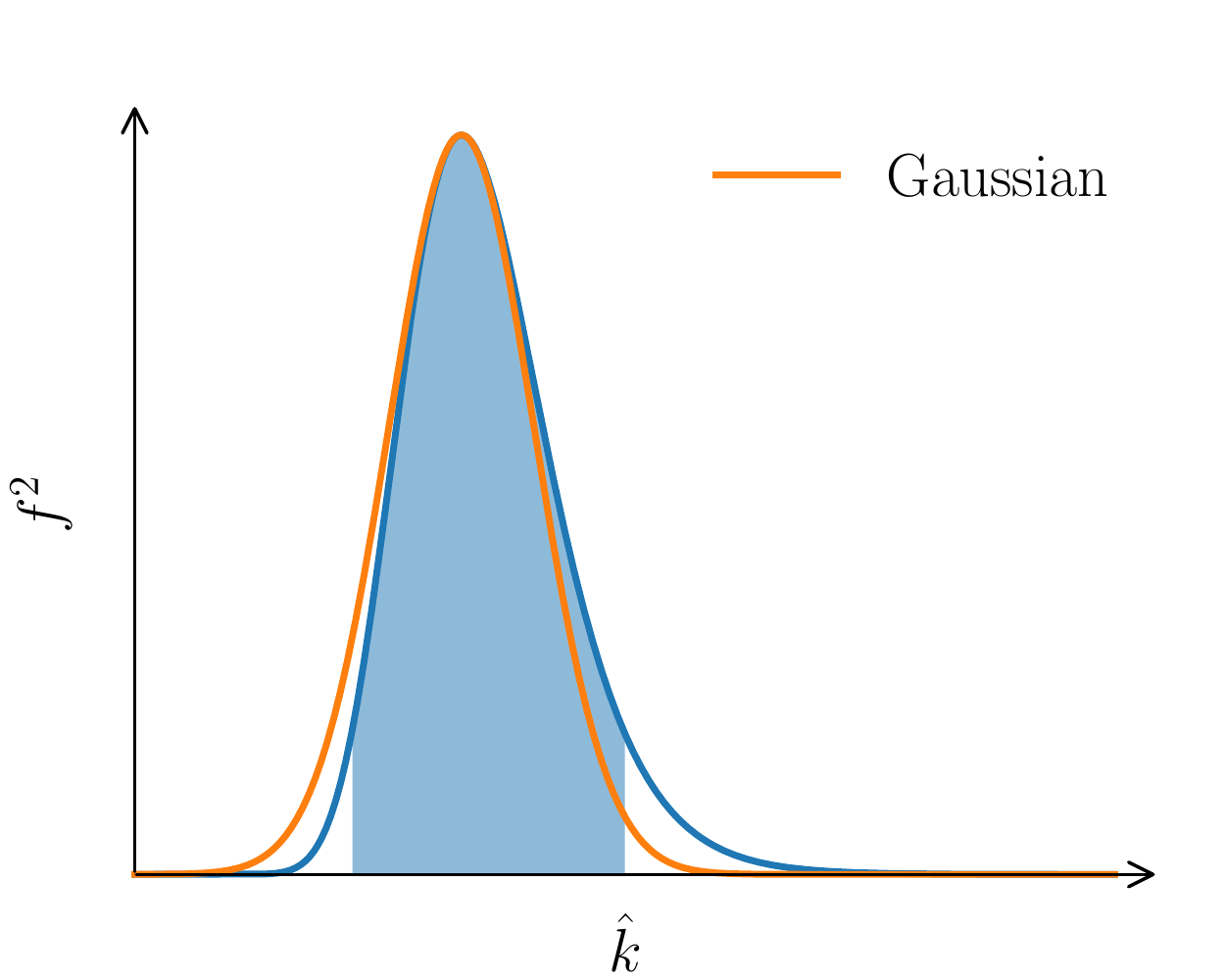}
\par\end{centering}
\caption{An illustration of the Gaussian integral approximation in Eq.~(\ref{eq:B2}).
Here, the blue curve is the function $f^{2}$, and the shaded area
is the integral of $f^{2}$ from $\hat{k}_{d}/\xi$ to $\hat{k}_{d}\xi$,
where $\xi=1.5$ is assumed. The orange curve is the Gaussian approximation
to the function $f^{2}$, and the domain of integration of the Gaussian
approximation is extended to $(-\infty,\infty)$.\label{fig:Gaussian}}

\end{figure}

Equation (\ref{eq:dominant_wavenumber}) gives the relation between
$\hat{k}_{d}$ and $\hat{a}_{d}$ but we still need to determine the
latter. This final step is accomplished by using the disruption condition
$\hat{w}=\hat{\delta}$. From Eqs.~(\ref{eq:island_width}) and (\ref{eq:inner_general}),
the disruption condition is equivalent to 
\begin{equation}
\frac{\tilde{B}^{2}}{B_{0}^{2}}=\frac{\hat{\gamma}_{d}}{16S},\label{eq:disruption-3-1-1}
\end{equation}
where the magnetic fluctuation $\tilde{B}$ is calculated by integrating
the fluctuation spectrum $f$ over a neighborhood of the dominant
wavenumber as prescribed in Eq.~(\ref{eq:amplitude}). Because the
fluctuation spectrum $f$ is localized around the dominant wavenumber,
we can first expand the exponent in Eq.~(\ref{eq:solution}) near
$\hat{k}_{d}$, then approximate the integration by a Gaussian integral.
Expanding $g$ in the neighborhood of $\hat{k}_{d}$ yields 
\begin{equation}
g\simeq g(\varphi_{d})+\frac{8}{25}g''(\varphi_{d})(\varphi_{d}/\hat{k}_{d})^{2}(\hat{k}-\hat{k}_{d})^{2}.\label{eq:g_expansion}
\end{equation}
Hence, in the neighborhood of the dominant mode, 
\begin{equation}
f(\hat{k})\simeq A\exp\left[\frac{8}{25}h(\hat{a}_{d})\,g''(\varphi_{d})(\varphi_{d}/\hat{k}_{d})^{2}(\hat{k}-\hat{k}_{d})^{2}\right],\label{eq:f_approx}
\end{equation}
where 
\begin{equation}
A\equiv f_{0}\left(\hat{k}_{d}e^{\hat{t}_{d}}\right)\exp\left[h(\hat{a}_{d})\,g(\varphi_{d})-\frac{\hat{t}_{d}}{2}\right]\label{eq:A}
\end{equation}
and $\hat{t}_{d}$ is the disruption time. Now we can evaluate the
magnetic fluctuation $\tilde{B}$ by using Eq.~(\ref{eq:f_approx})
and extending the domain of integration to $(-\infty,\infty)$ {[}see
Fig.~\ref{fig:Gaussian} for an illustration of this approximation{]},
yielding
\begin{align}
\frac{\tilde{B}^{2}}{B_{0}^{2}} & =\frac{1}{\pi}\int_{\hat{k}_{d}/\xi}^{\hat{k}_{d}\xi}f^{2}d\hat{k}\nonumber \\
 & \simeq\frac{A^{2}}{\pi}\int_{-\infty}^{\infty}\exp\left[\frac{16}{25}h(\hat{a}_{d})\,g''(\varphi_{d})(\varphi_{d}/\hat{k}_{d})^{2}(\hat{k}-\hat{k}_{d})^{2}\right]d\hat{k}\nonumber \\
 & =\frac{5A^{2}}{4\sqrt{\pi}}\frac{\hat{k}_{d}}{\varphi_{d}}\left(h(\hat{a}_{d})\left|g''(\varphi_{d})\right|\right)^{-1/2}.\label{eq:B2}
\end{align}
Using Eq.~(\ref{eq:B2}) and Eq.~(\ref{eq:dominant_gamma}) in the
disruption condition (\ref{eq:disruption-3-1-1}) yields the condition
to determine the disruption width $\hat{a}_{d}$:
\begin{gather}
\left[f_{0}\left(c_{k}S^{-1/4}\hat{a}_{0}^{\hat{\tau}}\hat{a}_{d}^{-5/4-\hat{\tau}}\right)\right]^{2}\exp\left[C_{\Gamma}^{5/8}g(\varphi_{d})\hat{\tau}_{\sharp}S^{-1/2}\hat{a}_{d}^{-3/2}\right]\nonumber \\
=\frac{\sqrt{\pi}}{20\sqrt{2}}\frac{c_{\gamma}\varphi_{d}}{c_{k}}C_{\Gamma}^{5/16}\left|g''(\varphi_{d})\right|^{1/2}\hat{\tau}_{\sharp}^{1/2}S^{-3/2}\hat{a}_{0}^{\hat{\tau}}\hat{a}_{d}^{-1-\hat{\tau}}.\label{eq:disruption}
\end{gather}
In deriving Eq.~(\ref{eq:disruption}), we have used Eq.~(\ref{eq:dominant_wavenumber})
for $\hat{k}_{d}$, Eq.~(\ref{eq:h}) for $h$, Eq.~(\ref{eq:A})
for $A$, and the relation $e^{\hat{t}_{d}}=(\hat{a}_{0}/\hat{a}_{d})^{\hat{\tau}}.$

Equation (\ref{eq:disruption}) is the fundamental equation to determine
the disruption current sheet width $\hat{a}_{d}$ for a general initial
condition $f_{0}(\hat{k}_{0})$; its validity only requires that $f_{0}(\hat{k}_{0})$
is slowly varying such that the dominant wavenumber in the spectrum,
given by Eq.~(\ref{eq:solution}), is determined by the peak in the
exponent. However, Eq.~(\ref{eq:disruption}) is difficult to solve
for a general $f_{0}(\hat{k}_{0})$ without resorting to numerical
methods. To make further progress, we assume that the initial condition
takes a power-law form $f_{0}(\hat{k}_{0})=\epsilon\hat{k}_{0}^{-\chi}$.
Then Eq.~(\ref{eq:disruption}) becomes 
\begin{multline}
\exp\left[C_{\Gamma}^{5/8}g\left(\varphi_{d}\right)\hat{\tau}_{\sharp}S^{-1/2}\hat{a}_{d}^{-3/2}\right]\\
=c_{\chi}\hat{\tau}_{\sharp}^{1/2}\epsilon^{-2}S^{-(\chi+3)/2}\hat{a}_{0}^{(2\chi+1)\hat{\tau}}\hat{a}_{d}^{-1-\hat{\tau}-5\chi/2-2\chi\hat{\tau}},\label{eq:disruption1}
\end{multline}
where 
\begin{equation}
c_{\chi}=\frac{\sqrt{\pi}}{20\sqrt{2}}\frac{c_{\gamma}\varphi_{d}}{c_{k}^{1-2\chi}}C_{\Gamma}^{5/16}\left|g''(\varphi_{d})\right|^{1/2}.\label{eq:c_chi}
\end{equation}
Equation (\ref{eq:disruption1}) can be solved for $\hat{a}_{d}$
in terms of the lower branch $W_{-1}$ of the Lambert $W$ function.\citep{CorlessGHJK1996}
The solution is
\begin{equation}
\hat{a}_{d}=c_{a}\hat{\tau}_{\sharp}^{2/3}S^{-1/3}\left[-\frac{1}{\theta}W_{-1}\left(\Xi\right)\right]^{-2/3},\label{eq:ad}
\end{equation}
where 
\begin{equation}
c_{a}\equiv\left[g\left(\varphi_{d}\right)C_{\Gamma}^{5/8}\right]^{2/3},\label{eq:ca}
\end{equation}
\begin{equation}
\theta\equiv\frac{3}{\left(4\chi+2\right)\hat{\tau}+5\chi+2},\label{eq:theta}
\end{equation}
and 
\begin{equation}
\Xi\equiv-\theta c_{a}^{3/2}\left(c_{\chi}\hat{a}_{0}^{(2\chi+1)\hat{\tau}}\right)^{-\theta}\epsilon^{2\theta}\hat{\tau}_{\sharp}^{1-\theta/2}S^{(\theta\chi+3\theta-1)/2}.\label{eq:Xi}
\end{equation}

An \emph{a posteriori} consistency check for the Gaussian integral
approximation in Eq.~(\ref{eq:B2}) is in order. Note that $g(\varphi_{d})$
and $g''(\varphi_{d})$ are typically $O(1)$ quantities; therefore,
the validity of the Gaussian integral approximation requires the condition
$h(\hat{a}_{d})\gg1$ being satisfied. Using Eq.~(\ref{eq:ad}) in
Eq.~(\ref{eq:h}), the condition is
\begin{equation}
h(\hat{a}_{d})=-\frac{C_{\Gamma}^{5/8}c_{a}^{-3/2}}{2\theta}W_{-1}\left(\Xi\right)\gg1.\label{eq:self-consistency}
\end{equation}
This condition is satisfied only if $\left|\Xi\right|\ll1$. Therefore,
the noise amplitude $\epsilon$ must be sufficiently small for the
approximation to be valid.

Because the condition $\left|\Xi\right|\ll1$ is satisfied, we can
use the asymptotic expansion of $W_{-1}(z)$ as $z\to0^{-}$, i.e.,

\begin{equation}
W_{-1}(z)=\log(-z)-\log\left(-\log(-z)\right)+o(1)\label{eq:W_function_expansion}
\end{equation}
 to obtain the leading order approximation for $\hat{a}_{d}$: 
\begin{equation}
\hat{a}_{d}\simeq c_{a}\hat{\tau}_{\sharp}^{2/3}S^{-1/3}\left[\log\left(\dfrac{\hat{a}_{0}^{(2\chi+1)\hat{\tau}}\hat{\tau}_{\sharp}^{1/2-1/\theta}}{\epsilon^{2}S^{(3+\chi-1/\theta)/2}}\right)\right]^{-2/3}.\label{eq:ad_approx}
\end{equation}
Here, we have ignored an $O(1)$ factor $c_{\chi}(\theta c_{a}^{3/2})^{-1/\theta}$
within the logarithm, which can be attributed to the next order correction
of the form $\log(\log(\ldots))$.

Now we have obtained the current sheet width $\hat{a}_{d}$ when disruption
occurs. The disruption time $\hat{t}_{d}$ can be obtained by $\hat{t}_{d}=\hat{\tau}\log\left(\hat{a}_{0}/\hat{a}_{d}\right)$.
The dominant wavenumber $\hat{k}_{d}$ and the linear growth rate
$\hat{\gamma}_{d}$ can be obtained by substituting $\hat{a}_{d}$
into Eqs.~(\ref{eq:dominant_wavenumber}) and (\ref{eq:dominant_gamma}).

\subsection{Case II --- System Noise \label{subsec:Case-II}}

The calculation in Section \ref{subsec:Case-I} assumes that an initial
perturbation is applied at $\hat{t}=0$. If the current sheet width
is sufficiently broad such that the linear growth rate $\hat{\gamma}(\hat{k})<1/2$,
then the perturbation amplitude at wavenumber $\hat{k}$ decreases
in time because of the advection loss caused by the reconnection outflow
{[}Eq.~(\ref{eq:model-normalized}){]}. If the linear growth rate
never rises above $1/2$, then the perturbation amplitude will decrease
monotonically in time and asymptotically approach zero as $\hat{t}\to\infty$.
However, since noise is present in any natural system, there will
always be some fluctuations in the system. Noise can be introduced
into the model by explicitly adding a source term or by setting a
lower bound (as a function of $\hat{k}$) to the fluctuation amplitude.
Here we adopt the second approach: Instead of an initial perturbation,
we now use $f_{0}(\hat{k})$ to describe the system noise, i.e., $f_{0}(\hat{k})$
represents the lower bound of $f(\hat{k}$). In this Section, we consider
the situation where the system noise provides the seed for the plasmoid
instability.

Because $\hat{\gamma}$ must be greater than $1/2$ for a mode to
grow when we integrate along a characteristic in Eq.~(\ref{eq:solution-3}),
the starting time $\hat{t}_{0}$ is determined by the condition $\hat{\gamma}=1/2$.
For neighboring modes of the dominant mode at the disruption time,
it can be shown \emph{a posteriori }that those modes start to grow
when they are in the small-$\Delta'$ regime.\footnote{This can be seen as follows. Note that the dominant wavenumber $\hat{k}_{d}$
scales in the same way as the fastest-growing wavenumber $\hat{k}_{\text{max}}\sim S^{-1/4}\hat{a}_{d}{}^{-5/4}$
at the disruption time, where the latter is at the transition between
the small-$\Delta'$ and the large-$\Delta'$ regimes. Because $\hat{k}=\hat{k}_{d}e^{\hat{t}_{d}-\hat{t}}$
along a characteristic and $\hat{a}\propto e^{-\hat{t}/\hat{\tau}}$,
the wavenumber $\hat{k}$ is larger than $\hat{k}_{\max}\sim S^{-1/4}\hat{a}_{d}e^{-5(\hat{t}_{d}-\hat{t})/4\hat{\tau}}$
at an earlier time $\hat{t}$ with $\hat{t}_{d}-\hat{t}>O(1)$, which
places the mode in the small-$\Delta'$ regime.} Using the relations $\hat{k}_{0}=\hat{k}e^{\hat{t}-\hat{t}_{0}}$
and $\hat{a}(\hat{t}_{0})=\hat{a}e^{(\hat{t}-\hat{t}_{0})/\hat{\tau}}$
in the condition using the small-$\Delta'$ dispersion relation
\begin{equation}
\hat{\gamma}_{s}(\hat{k}_{0},\hat{t}_{0})=C_{\Gamma}\hat{a}(\hat{t}_{0})^{-2}S^{-3/5}\hat{k}_{0}^{-2/5}=\frac{1}{2}\label{eq:start-condition}
\end{equation}
yields 
\begin{equation}
\hat{t}-\hat{t}_{0}=\frac{\hat{\tau}_{\sharp}}{2}\log\left(2C_{\Gamma}S^{-3/5}\hat{a}^{-2}\hat{k}^{-2/5}\right);\label{eq:start-time}
\end{equation}
hence, the wavenumber $\hat{k}_{0}$ when the mode starts to grow
is 
\begin{equation}
\hat{k}_{0}=\hat{k}e^{\hat{t}-\hat{t}_{0}}=\left(2C_{\Gamma}S^{-3/5}\hat{a}^{-2}\right)^{\hat{\tau}_{\sharp}/2}\hat{k}^{1-\hat{\tau}_{\sharp}/5}.\label{eq:k0}
\end{equation}

Following the same calculation leading to Eq.~(\ref{eq:solution}),
the solution is 
\begin{align}
f(\hat{k},\hat{t}) & \simeq f_{0}(\hat{k}_{0})\exp\left[h\,g(\varphi;\hat{\tau})-\frac{\hat{t}-\hat{t}_{0}}{2}\right],\label{eq:solution-noise-only}
\end{align}
with Eq.~(\ref{eq:start-time}), Eq.~(\ref{eq:k0}), $h=\hat{\tau}_{\sharp}C_{\Gamma}^{5/8}S^{-1/2}\hat{a}^{-3/2}/2$,
and $\varphi=C_{\Gamma}^{-3/4}S^{1/5}\hat{k}^{4/5}\hat{a}$ being substituted into
the right-hand-side.

Equations (\ref{eq:dominant_wavenumber}) -- (\ref{eq:g_expansion})
remain valid, but $A$ in Eqs.~(\ref{eq:f_approx}) -- (\ref{eq:B2})
must be replaced by
\begin{align}
\tilde{A}= & f_{0}\text{\ensuremath{\left(\left(2C_{\Gamma}S^{-3/5}\hat{a}_{d}^{-2}\right)^{\hat{\tau}_{\sharp}/2}\hat{k}_{d}^{1-\hat{\tau}_{\sharp}/5}\right)}}\nonumber \\
 & \text{\ensuremath{\left(2C_{\Gamma}S^{-3/5}\hat{a}_{d}^{-2}\hat{k}_{d}^{-2/5}\right)}}^{-\hat{\tau}_{\sharp}/4}\exp\left[h(\hat{a}_{d})\,g\left(\varphi_{d}\right)\right].\label{eq:A1}
\end{align}
The disruption condition (\ref{eq:disruption-3-1-1}) yields the equation
to determine $\hat{a}_{d}$:
\begin{multline}
\left[f_{0}\text{\ensuremath{\left(c_{k}{}^{1-\hat{\tau}_{\sharp}/5}\left(2C_{\Gamma}\right)^{\hat{\tau}_{\sharp}/2}S^{-\hat{\tau}_{\sharp}/4-1/4}\hat{a}_{d}^{-3\hat{\tau}_{\sharp}/4-5/4}\right)}}\right]^{2}\\
\exp\left[\hat{\tau}_{\sharp}C_{\Gamma}^{5/8}S^{-1/2}\hat{a}_{d}^{-3/2}g\left(\varphi_{d}\right)\right]\\
=\frac{\sqrt{\pi}c_{\gamma}\varphi_{d}}{20\sqrt{2}}\text{\ensuremath{c_{k}}}^{-1-\hat{\tau}_{\sharp}/5}C_{\Gamma}^{5/16}\left|g''(\varphi_{d})\right|^{1/2}\\
\text{\ensuremath{\left(2C_{\Gamma}\right)}}^{\hat{\tau}_{\sharp}/2}\hat{\tau}_{\sharp}^{1/2}S^{-3/2-\hat{\tau}_{\sharp}/4}\hat{a}_{d}^{-1-3\hat{\tau}_{\sharp}/4}.\label{eq:disruption2}
\end{multline}
If we assume a power-law system noise $f_{0}(\hat{k})=\epsilon\hat{k}^{-\chi}$,
Eq.~(\ref{eq:disruption2}) becomes
\begin{multline}
\exp\left[C_{\Gamma}^{5/8}g(\varphi_{d})\hat{\tau}_{\sharp}S^{-1/2}\hat{a}_{d}^{-3/2}\right]\\
=\tilde{c}_{\chi}\epsilon^{-2}\hat{\tau}_{\sharp}^{1/2}S^{-(3+\chi)/2-(2\chi+1)\hat{\tau}_{\sharp}/4}\\
\hat{a}_{d}^{-1-5\chi/2-(3+6\chi)\hat{\tau}_{\sharp}/4},\label{eq:disruption3}
\end{multline}
where 
\begin{equation}
\tilde{c}_{\chi}\equiv c_{\chi}\text{\ensuremath{\left(2C_{\Gamma}c_{k}^{-2/5}\right)}}^{(2\chi+1)\hat{\tau}_{\sharp}/2}.\label{eq:c_chi_1}
\end{equation}
The solution for $\hat{a}_{d}$ can be expressed in terms of the lower-branch
Lambert function $W_{-1}$ as
\begin{equation}
\hat{a}_{d}=c_{a}\hat{\tau}_{\sharp}^{2/3}S^{-1/3}\left[-\frac{1}{\tilde{\theta}}W_{-1}\left(\tilde{\Xi}\right)\right]^{-2/3},\label{eq:ad1}
\end{equation}
where
\begin{equation}
\tilde{\theta}\equiv\frac{6}{4+10\chi+(6\chi+3)\hat{\tau}_{\sharp}},\label{eq:theta1}
\end{equation}
\begin{align}
\tilde{\Xi}\equiv & -\tilde{\theta}c_{a}^{3/2}\tilde{c}_{\chi}^{-\tilde{\theta}}\epsilon^{2\tilde{\theta}}\hat{\tau}_{\sharp}^{1-\tilde{\theta}/2}S^{(\tilde{\theta}\chi+3\tilde{\theta}-1+\tilde{\theta}(\chi+1/2)\hat{\tau}_{\sharp})/2},\label{eq:Xi-1}
\end{align}
and $c_{a}$ is defined in Eq.~(\ref{eq:ca}). Using the asymptotic
expansion (\ref{eq:W_function_expansion}), the leading order approximation
of $\hat{a}_{d}$ is
\begin{equation}
\hat{a}_{d}\simeq c_{a}\hat{\tau}_{\sharp}^{2/3}S^{-1/3}\left[\log\left(\dfrac{\hat{\tau}_{\sharp}^{1/2-1/\tilde{\theta}}}{\epsilon^{2}S^{(\chi+3+(\chi+1/2)\hat{\tau}_{\sharp}-1/\tilde{\theta})/2}}\right)\right]^{-2/3}.\label{eq:ad1_approx}
\end{equation}
Here, we again ignore an $O(1)$ factor within the logarithm, which
can be attributed to the next order correction of the form $\log(\log(\ldots))$.

\section{Discussions\label{sec:Discussions}}

\subsection{Effects of The Reconnection Outflow}

Our phenomenological model includes the effects of the reconnection
outflow. It is instructive to investigate how the outflow affects
the scaling relations. The effects of the outflow can be turned off
by setting $v_{o}$ to zero in Eq.~(\ref{eq:model}). While we have
repeated the calculation for cases without the outflow, here we present
an easier way that yields the same results. Neglecting the effects
of the outflow can be formally achieved by assuming that the current
sheet thins down almost instantaneous, i.e., the current sheet thinning
time scale $\tau\ll\tau_{A}$, such that the mode-stretching and advective
loss due to the outflow have no time to take effect. Taking the limit
$\hat{\tau}=\tau/\tau_{A}\to0$ is thus equivalent to neglecting the
effects of the outflow. In this limit, the scaling relations obtained
in Sections \ref{subsec:Case-I} and \ref{subsec:Case-II} reduce
to the same results; i.e., it makes no difference whether the plasmoid
instability evolves from an initial noise or a system noise if the
effects of the outflow are ignored.

In the $\hat{\tau}\to0$ limit, the two time scales $\hat{\tau}_{*}$
and $\hat{\tau}_{\sharp}$ become identical to $\hat{\tau}$. The
variables $\theta$ and $\Xi$ become
\begin{equation}
\tilde{\tilde{\theta}}=\frac{3}{5\chi+2}\label{eq:theta2}
\end{equation}
and
\begin{equation}
\tilde{\tilde{\Xi}}=-\theta c_{a}^{3/2}c_{\chi}^{-\theta}\epsilon^{2\theta}\hat{\tau}^{1-\theta/2}S^{(\theta\chi+3\theta-1)/2}.\label{eq:Xi2}
\end{equation}
The disruption current sheet width is given by
\begin{equation}
\hat{a}_{d}=c_{a}\hat{\tau}_{\sharp}^{2/3}S^{-1/3}\left[-\frac{1}{\tilde{\tilde{\theta}}}W_{-1}\left(\tilde{\tilde{\Xi}}\right)\right]^{-2/3}\label{eq:ad4}
\end{equation}
with the leading order approximation 
\begin{align}
\hat{a}_{d} & \simeq c_{a}\hat{\tau}^{2/3}S^{-1/3}\left[\log\left(\dfrac{\hat{\tau}^{1/2-1/\tilde{\tilde{\theta}}}}{\epsilon^{2}S^{(3+\chi-1/\tilde{\tilde{\theta}})/2}}\right)\right]^{-2/3}\nonumber \\
 & =c_{a}\hat{\tau}^{2/3}S^{-1/3}\left[\log\left(\epsilon^{-2}\hat{\tau}^{-(1+10\chi)/6}S^{(2\chi-7)/6}\right)\right]^{-2/3}.\label{eq:ad4_approx}
\end{align}
Therefore, ignoring the outflow changes the time scale from $\hat{\tau}_{\sharp}$
to $\hat{\tau}$ in the scaling relation of $\hat{a}_{d}$; the power
indices of $\hat{\tau}$ and $S$ within the logarithmic factor are
also different. Otherwise, the scaling relation (\ref{eq:ad4_approx})
is similar to Eqs. (\ref{eq:ad_approx}) and (\ref{eq:ad1_approx}).

\subsection{Comparison with the Scaling Relations of Comisso et al.}

Now we are in a position to compare the analytic scaling relations
in this work with that obtained previously by Comisso \emph{et al.},\citep{ComissoLHB2016,ComissoLHB2017}
because the latter also ignore the effects of outflow. However, as
Comisso \emph{et al.} used a different way to describe the fluctuations,
we need a translation of notations before a comparison can be made.

Comisso \emph{et al.} describe the fluctuations by the island size
$\hat{w}=2(\hat{\psi}\hat{a})^{1/2}$ as a function of $\hat{k}$
and $\hat{t}$, where $\hat{\psi}$ is the magnetic flux of the island.
They propose a ``principle of least time'' that determines the dominant
mode to disrupt the current sheet by the wavenumber that satisfies
the condition $\hat{w}(\hat{k},\hat{t})=\hat{\delta}(\hat{k},\hat{t})$
with the shortest time. For an initial condition $\hat{\psi}_{0}(\hat{k})=\epsilon\hat{k}^{-\alpha}$,
the leading order approximation of the current sheet width at disruption
is given by Eq.~(26) of Ref.~{[}\onlinecite{ComissoLHB2017}{]}:
\begin{equation}
\hat{a}_{d}\simeq c_{a}\hat{\tau}^{2/3}S^{-1/3}\left[\log\left(\epsilon^{-2}\hat{\tau}^{(2-5\alpha)/3}S^{(\alpha-4)/3}\right)\right]^{-2/3}.\label{eq:ad_comisso}
\end{equation}

The present work describes the fluctuations by a spectrum $f(\hat{k},\hat{t})$
of the magnetic field. To compare Eq.~(\ref{eq:ad_comisso}) with
Eq.~(\ref{eq:ad4_approx}), we need to find a correspondence between
$f$ and $\hat{\psi}$. This can be obtained by noting that the magnetic
field fluctuation $\tilde{B}\sim\hat{k}^{1/2}fB_{0}$ {[}from Eq.~(\ref{eq:amplitude}){]}
and $\tilde{B}\sim\hat{k}\hat{\psi}B_{0}$ {[}from Eq.~(\ref{eq:island_width})
and the relation $\hat{w}=2(\hat{\psi}\hat{a})^{1/2}${]}; hence,
the correspondence is $f\sim\hat{k}^{1/2}\hat{\psi}$. For the assumed
power-law initial perturbations $f=\epsilon\hat{k}^{-\chi}$ and $\hat{\psi}=\epsilon\hat{k}^{-\alpha}$,
we have the correspondence $\alpha\leftrightarrow\chi+1/2$. Substituting
$\alpha\rightarrow\chi+1/2$ into Eq.~(\ref{eq:ad4_approx}), we
recover the scaling relation (\ref{eq:ad_comisso}). Therefore, we
conclude that when the effects of the outflow are ignored, the phenomenological
model of this work yields the same scaling relations as the relations
obtained using the principle of least time. However, the agreement
is only up to the leading order approximation. It can be shown that
the two approaches give different results if the next order corrections
of the Lambert $W$ function of the form $\log(\log(\ldots))$ are
included.

\subsection{Comparison with Numerical Solutions of the Model Equation\label{subsec:Comparison-with-Numerical}}

\begin{table}
\begin{tabular}{|c|c|c|c|c|}
\hline 
 & $\hat{\tau}=\log2,$ & $\hat{\tau}\to0,$ & $\hat{\tau}=\log2,$ & $\hat{\tau}=\log2,$\tabularnewline
 & $\chi=0$ & $\chi=0$ & $\chi=1$ & $\chi=3$\tabularnewline
\hline 
$c_{a}$ & 0.873 & 0.838 & 0.873 & 0.873\tabularnewline
\hline 
$c_{k}$ & 0.563 & 0.705 & 0.563 & 0.563\tabularnewline
\hline 
$c_{\gamma}$ & 0.538 & 0.576 & 0.538 & 0.538\tabularnewline
\hline 
$c_{\chi}$ & 0.0349 & 0.0313 & 0.0111 & 0.00111\tabularnewline
\hline 
\end{tabular}

\caption{The values of constants $c_{a}$, $c_{k}$, $c_{\gamma}$, and $c_{\chi}$
for different settings reported in this work. \label{tab:constants}}
\end{table}
\begin{figure*}
\begin{centering}
\includegraphics[scale=0.7]{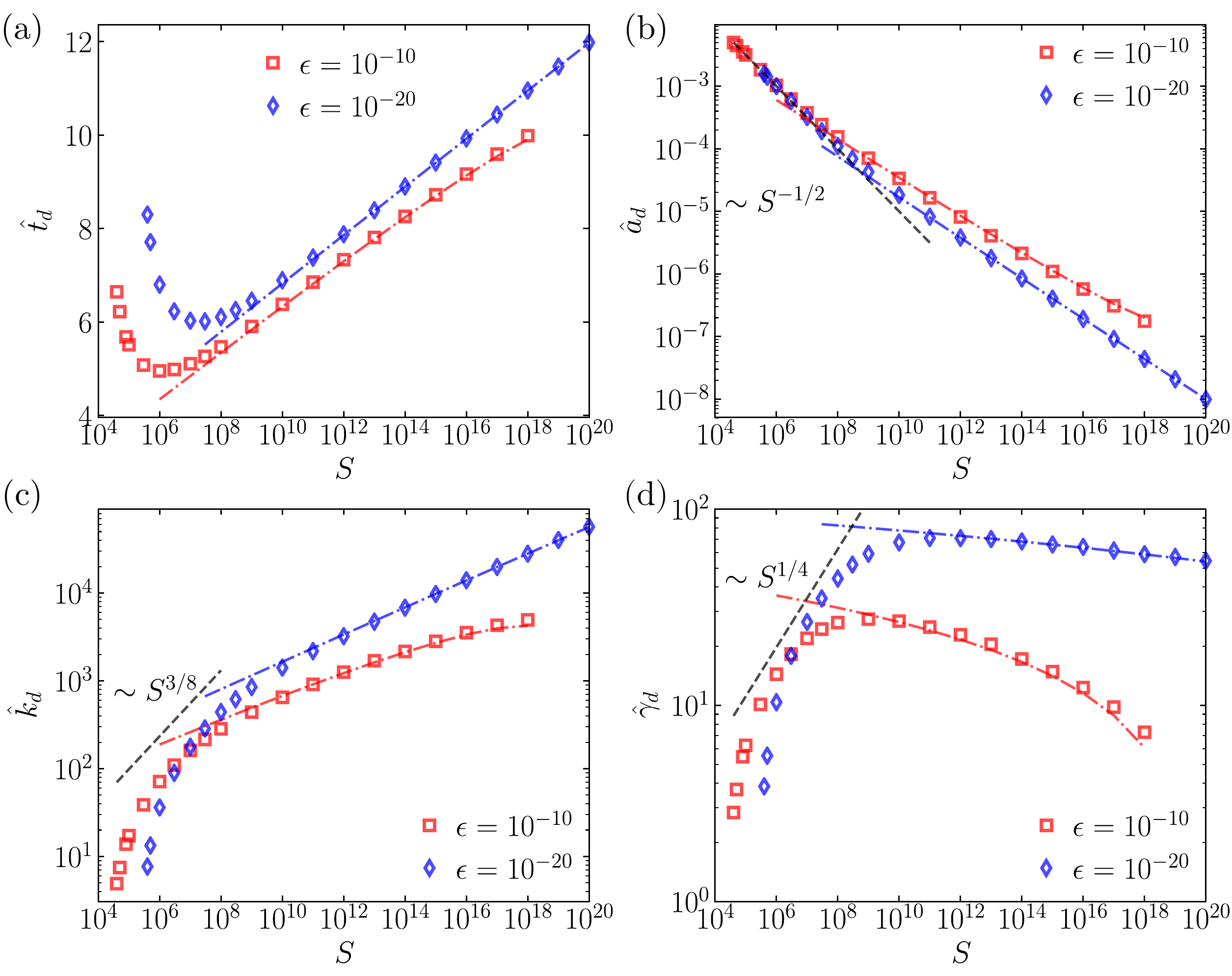}
\par\end{centering}
\caption{Scalings of $\hat{t}_{d}$, $\hat{a}_{d}$, $\hat{k}_{d}$, and $\hat{\gamma}_{d}$
with respect to the Lundquist number $S$. Here, a constant initial
noise spectrum $f_{0}(\hat{k})=\epsilon$ is assumed. The symbols
are results obtained by numerical solutions of the model equation,
whereas the dash-dotted lines are analytic scaling relations. The
black dashed lines are scalings based on the fastest growing mode
of a Sweet-Parker current sheet. \label{fig:scaling1}}
\end{figure*}
\begin{figure}
\begin{centering}
\includegraphics[scale=0.65]{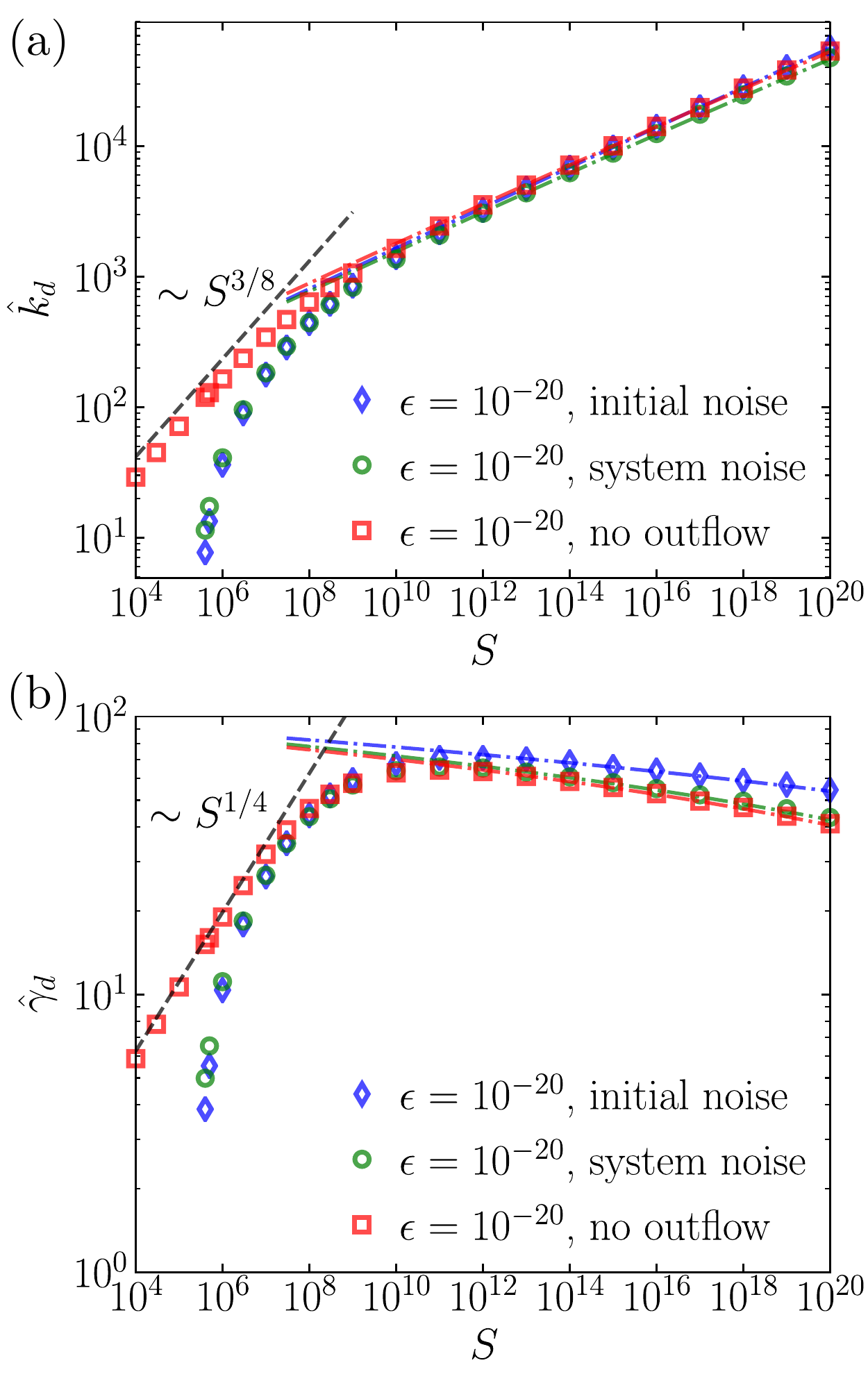}\caption{Scalings of $\hat{k}_{d}$ and $\hat{\gamma}_{d}$ with respect to
the Lundquist number $S$. Here, a constant noise spectrum $f_{0}(\hat{k})=\epsilon$
is assumed. \label{fig:scaling2}}
\par\end{centering}
\end{figure}
\begin{figure}
\begin{centering}
\includegraphics[scale=0.65]{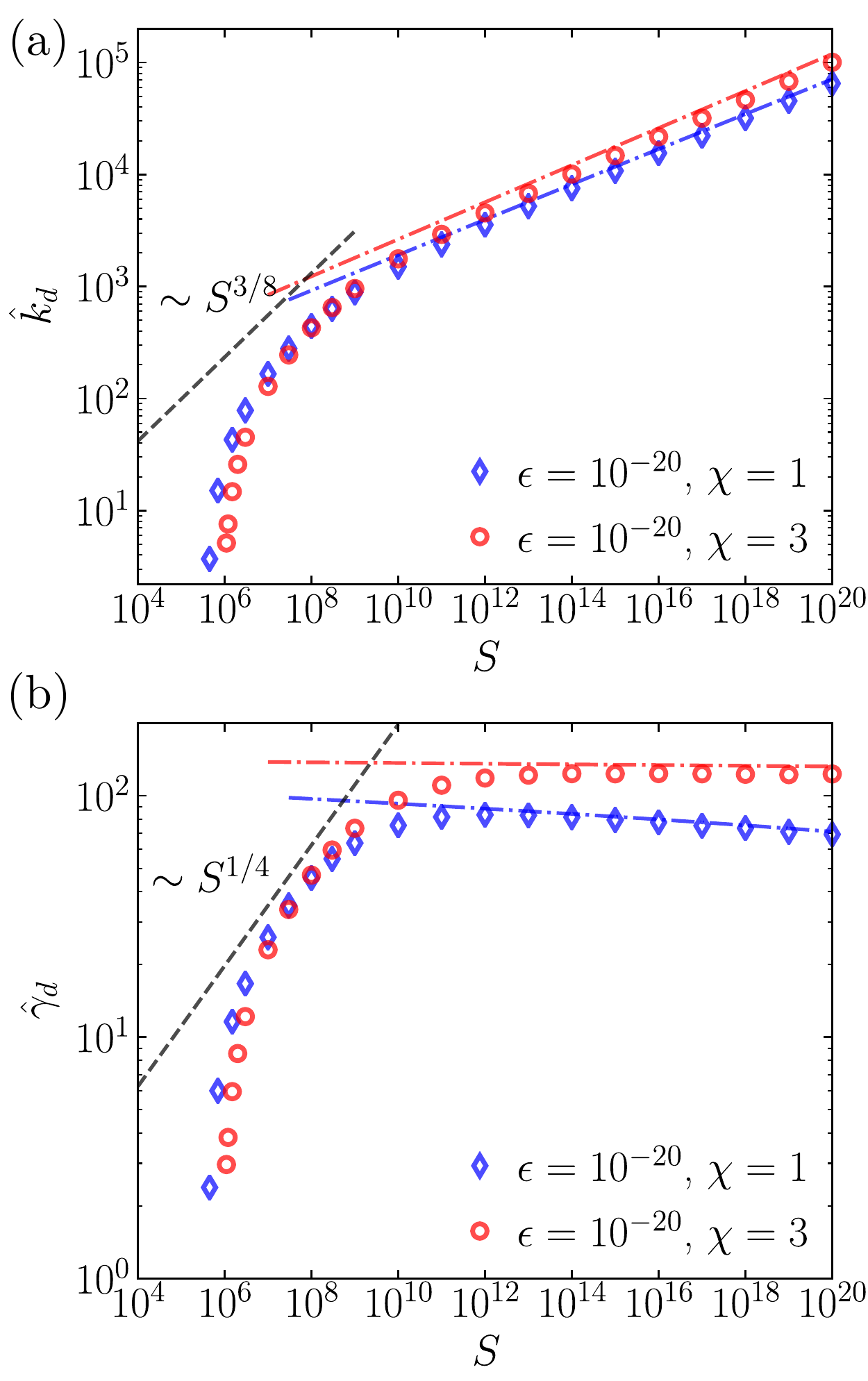}
\par\end{centering}
\caption{Scalings of $\hat{k}_{d}$ and $\hat{\gamma}_{d}$ with respect to
the Lundquist number $S$. Here, a power-law system noise $f_{0}(\hat{k})=\epsilon\hat{k}^{-\chi}$
is assumed.\label{fig:scaling3}}
\end{figure}

We have employed several approximations when deriving the analytic
scaling relations. These approximations include: (i) ignoring the
asymptotic current sheet width $a_{\infty}$ in Eq.~(\ref{eq:a-solution});
(ii) using an approximate linear growth rate Eq.~(\ref{eq:gamma})
that ignores the $\hat{k}\hat{a}=1$ stability threshold; (iii) ignoring
the contribution from the lower bound $\varphi_{0}$ of the integral
in Eq.~(\ref{eq:gamma_int_0}); (iv) assuming the initial or system
noise spectrum $f_{0}(\hat{k})$ to be slowly varying compared to
the exponential factor in Eq.~(\ref{eq:solution}); and (v) replacing
the integral in (\ref{eq:B2}) by a Gaussian integral and pushing
the bounds to $\pm\infty$. Among these approximations, (iv) is an
assumption that must be satisfied by the noise spectrum $f(\hat{k}_{0})$;
the other approximations can be justified \emph{a posteriori} in the
limit of high-$S$ and low-noise.

We now test the analytic scaling relations with the results from numerical
solutions of the model equation. For the numerical solutions, we do
not employ any approximation except the linear growth rate, where
Eq.~(\ref{eq:uniform_approx}) with $\zeta=3/2$ is used. Using the
approximate linear growth rate is not a significant compromise because
the expression is nearly exact; note that the analytic calculation
employs a further simplified expression for the linear growth rate,
Eq.~(\ref{eq:gamma}).

To fix ideas, we employ the same setup as in Ref.~{[}\onlinecite{HuangCB2017}{]},
i.e., with $\hat{a}_{0}=1/\pi$ and $\hat{\tau}=\log2$. We vary the
parameters $\epsilon$, $\chi$, and $S$ for cases with either an
initial noise or a system noise, with or without the outflow. We integrate
the normalized model equation (\ref{eq:model-normalized}) forward
in time along characteristics. Because the wavenumber $\hat{k}$ decreases
exponentially in time along a characteristic, it is convenient to
set the mesh in the wavenumber $\hat{k}$ equispaced with respect
$\log\hat{k}$ and set the time step $\Delta\hat{t}=\Delta\log\hat{k}$.
In this way, the wavenumber $\hat{k}$ shifts exactly one grid space
in one time step, which significantly simplifies the time integration
along characteristics. We perform the time integration with a second-order
trapezoidal scheme. For cases with system noise, in each time step
the solution $f(\hat{k})$ is compared with $f_{0}(\hat{k})$; we
set $f(\hat{k})$ to $f_{0}(\hat{k})$ for those $\hat{k}$ where
$f(\hat{k})<f_{0}(\hat{k})$. In each time step, we identify the dominant
wavenumber and calculate the plasmoid size $\hat{w}$ with Eq.~(\ref{eq:amplitude}),
where $\xi=1.5$ has been used. The time integration continues until
the disruption condition $\hat{\delta}=\hat{w}$ is met. Then we record
$\hat{a}_{d}$, $\hat{k}_{d}$, and $\hat{\gamma}_{d}$ at the disruption
time $\hat{t}_{d}$.

The analytic scaling relations involve constants $c_{a}$, $c_{k}$,
$c_{\gamma}$, and $c_{\chi}$. These constants depend on $\hat{\tau}$
because they depend on $\varphi_{d}$ that maximizes the function
$g(\varphi;\hat{\tau})$; additionally, $c_{a}$ and $c_{\chi}$ also
depend on $g(\varphi_{d})$ and $g''(\varphi_{d})$. For $\hat{\tau}=\log2$,
we numerically obtain $\varphi_{d}\approx0.655$, $g(\varphi_{d})\approx0.841$,
and $g''(\varphi_{d})\approx-0.820$. If the outflow is neglected,
i.e., taking the limit $\hat{\tau}\to0$, they become $\varphi_{d}\approx0.784$,
$g(\varphi_{d})\approx0.791$, and $g''(\varphi_{d})\approx-0.629$.
These values are used in Eqs.~(\ref{eq:ck}), (\ref{eq:cgamma}),
(\ref{eq:c_chi}), and (\ref{eq:ca}) to obtain the constants $c_{a}$,
$c_{k}$, $c_{\gamma}$, and $c_{\chi}$. The values of these constants
for different settings reported in this work are listed in Table \ref{tab:constants}.
In the following comparisons, we use the precise analytic scaling
relations involving the Lambert function, but the leading order approximate
scaling relations are almost as good.

We start with a setup having the initial noise $f_{0}(\hat{k})=\epsilon$.
The scalings for $\hat{t}_{d}$, $\hat{a}_{d}$, $\hat{k}_{d}$, and
$\hat{\gamma}_{d}$ with respect to $S$ are shown in Fig.~\ref{fig:scaling1}
for two cases $\epsilon=10^{-10}$ and $\epsilon=10^{-20}$. Here,
the symbols denote the scalings obtained from numerical solutions,
the dash-dotted lines are the analytic scalings, and the black dashed
lines are scalings based on the fastest growing mode of the Sweet-Parker
current sheet {[}i.e., by setting $\hat{a}=\hat{a}_{SP}=S^{-1/2}$
in Eqs.~(\ref{eq:kmax}) and (\ref{eq:gamma_max}){]}. We can see
that the analytic scalings accurately agree with the numerically results
in the high-$S$ regime. On the other hand, while the Sweet-Parker
scaling $\hat{a}_{SP}=S^{-1/2}$ agrees with $\hat{a}_{d}$ in the
low to moderately-high $S$ regimes, the Sweet-Parker-based scalings
for $\hat{k}_{\text{max}}$ and $\hat{\gamma}_{\text{max}}$ considerably
depart from the numerical results of $\hat{k}_{d}$ and $\hat{\gamma}_{d}$;
the difference is especially substantial when $S$ is low. The discrepancies
are due to the stretching effect of outflow that makes the wavelength
of the dominant mode significantly longer than that of the fastest
growing mode.

Next, we test the analytic scalings for the other two cases: (1) system
noise instead of initial noise and (2) ignoring the effects of outflow.
Here, we set $f_{0}(\hat{k})=\epsilon$ with $\epsilon=10^{-20}$.
We only present the results for $\hat{k}_{d}$ and $\hat{\gamma}_{d}$
since they provide more stringent tests than $\hat{t}_{d}$ and $\hat{a}_{d}$.
The scalings of $\hat{k}_{d}$ and $\hat{\gamma}_{d}$ with respect
to $S$ are shown in Fig.~\ref{fig:scaling2}; for comparison, we
also show results that have been presented in Fig.~\ref{fig:scaling1}
for cases of initial noise. We again see that the analytic scalings
are in excellent agreement with the numerical results in the high-$S$
regime. In that regime, the scalings of $\hat{k}_{d}$ are quite similar
for all three cases; on the other hand, ignoring the effect of the
outflow only slightly affects the scaling of $\hat{\gamma}_{s}$ compared
to the case with a system noise, whereas $\hat{\gamma}_{s}$ is higher
for the case with an initial noise. The linear growth rate $\hat{\gamma}_{s}$
is higher for the initial noise case because the noise decays during
the early period when all the modes are stable; therefore, the disruption
occurs at a later time compared to the other two cases. As the current
sheet is thinner at a later time, the linear growth rate is higher.
Note that the Sweet-Parker-based scalings agree quite well with numerical
results at the low-$S$ regime when the outflow is ignored, confirming
our earlier statement that the discrepancies are due to the effect
of the outflow.

The analytic scaling relations are derived under the assumption that
the noise spectrum $f_{0}(\hat{k})$ is slowly varying compared to
the exponential factor in Eq.~(\ref{eq:solution}). For the cases
we have tested so far, this assumption is trivially satisfied with
$f_{0}(\hat{k})=\epsilon$. Now we test the analytic scalings for
a power-law system noise $f_{0}(\hat{k})=\epsilon\hat{k}^{-\chi}$,
with $\epsilon=10^{-20}$. Figure \ref{fig:scaling3} shows the scalings
of $\hat{k}_{d}$ and $\hat{\gamma}_{d}$ for the cases $\chi=1$
and $\chi=3$. We can see that even for a relatively steep noise spectrum
with $\chi=3$, the analytical scalings remain quite accurate.

\section{Conclusion\label{sec:Conclusion}}

In conclusion, we have derived analytic scaling relations for the
phenomenological model of plasmoid-mediated disruption of evolving
current sheets, including the effects of the reconnection outflow.
The analytic scaling relations are derived for two scenarios, one
with an initial noise and another with a system noise. The former
is useful for comparison with numerical simulations or laboratory
experiments where an initial perturbation can be injected; the latter
is more suitable for describing the plasmoid instability in natural
systems. If the effects of outflow are neglected, and with a proper
translation of notations, the scalings obtained from the present model
agree with that from previous works based on a principle of least
time,\citep{ComissoLHB2016,ComissoLHB2017} up to the leading order
approximation.

The phenomenological model is valid in the linear regime of the plasmoid
instability, up to the point of current sheet disruption. The time
scale $t_{d}$ for current sheet disruption is typically on the order
of a few Alfv\'en time $\tau_{A}$. Here, we estimate the Alfv\'en
times for various systems of interest. For a large-scale coronal mass
ejection (CME), the post-CME current sheet can extend to a length
beyond several times of the solar radius $R_{\odot}\simeq7\times10^{8}\text{m}$;
the Alfv\'en speed can be estimated from the speeds of moving blobs
in the current sheet, typically in the range of $2\times10^{5}$ --
$10^{6}\text{m/s}$.\citep{GuoBH2013,LinMSRRZWL2015} Assuming $L\sim10^{9}\text{m}$
and $V_{A}\sim10^{6}\text{m/sec}$, the Alfv\'en time $\tau_{A}$
is on the order of $1000$ seconds. For ultraviolet bursts in the
solar transition region,\citep{InnesGHB2015,ChittaPYH2017} we estimate
the current sheet length $L\sim10^{6}\text{m}$ from forward modeling
\citep{PeterHCY2019} and the Alfv\'en speed $V_{A}\sim10^{5}\text{m/sec}$
from the Doppler broadening of emission line profiles, yielding $\tau_{A}\sim10$
seconds. For laboratory magnetic reconnection experiments,\citep{Jara-AlmonteJYYF2016,OlsonEG2016}
the Alfv\'en time $\tau_{A}$ is typically on the order of microseconds,
which is within the temporal resolution of in situ measurements.\footnote{H. Ji, private communication (2019)}
Thus, it may be possible to test some of the predictions from the
phenomenological model with future generations of laboratory experiments.\citep{JiCGGGHJKKRSY2018}

Although the present study focuses on resistive MHD, the methodology
is quite general. The analysis could be generalized to include other
effects such as viscosity, the Hall effect, or ambipolar diffusion
from partially ionized plasmas. It is also important to further test
the predictions of the phenomenological model with full resistive
MHD simulations, especially in the high-$S$ regime. While the present
analytic calculation is limited to the special class of evolving current
sheets where the upstream magnetic field remains constant, the phenomenological
model could also describe cases where the upstream magnetic field
evolves in time, e.g., as in Refs. {[}\onlinecite{ShayDSR2004}{]}
and {[}\onlinecite{CassakD2009}{]}. These directions will be pursued
in the future.
\begin{acknowledgments}
We thank the anonymous referee for many constructive suggestions.
This work is supported by the National Science Foundation, Grant Nos.~AGS-1338944
and AGS-1460169, the Department of Energy, Grant No.~DE-SC0016470,
and the National Aeronautics and Space Administration, Grant No.~80NSSC18K1285.
A. Bhattacharjee would like to acknowledge the hospitality of the
Flatiron Institute, supported by the Simons Foundation.
\end{acknowledgments}

\end{document}